%% file: infocom.tex
\documentclass[conference]{IEEEtran}
\IEEEoverridecommandlockouts
\usepackage{cite}
\usepackage{amsmath,amssymb,amsfonts}
\usepackage{algorithmic}
\usepackage{graphicx}
\usepackage{textcomp}
\usepackage{xcolor}
\usepackage{booktabs}
\usepackage{multirow}
\usepackage{tabularx}
\usepackage{subcaption}
\usepackage[ruled,linesnumbered,vlined]{algorithm2e}
\newcommand{\head}[1]{\noindent\textbf{#1}}
\newcommand{\sysname}{\textit{Chameleon}\xspace}

\newtheorem{Definition}{Definition}
\def\BibTeX{{\rm B\kern-.05em{\sc i\kern-.025em b}\kern-.08em
    T\kern-.1667em\lower.7ex\hbox{E}\kern-.125emX}}
\begin{document}

\title{\sysname: Adaptive Fault Tolerance for Distributed Training via Real-time Policy Selection}


\author{
{\normalfont\fontsize{12pt}{15pt}\selectfont 
Yuhang Zhou$^{1}$, 
Zhibin Wang$^{1,*}$\thanks{* Corresponding author: wzbwangzhibin@nju.edu.cn}, 
Peng Jiang$^{1}$, 
Haoran Xia$^{1}$, 
Junhe Lu$^{1}$, 
} \\ 
{\normalfont\fontsize{12pt}{15pt}\selectfont  
Qianyu Jiang$^{1}$, 
Rong Gu$^{1}$, 
Hengxi Xu$^{2}$, 
Xinjing Huang$^{2}$, 
Guanghuan Fang$^{2}$,
} \\
{\normalfont\fontsize{12pt}{15pt}\selectfont  
Zhiheng Hu$^{2}$,
Jingyi Zhang$^{2}$, 
Yongjin Cai$^{2}$, 
Jian He$^{2}$,
Chen Tian$^{1}$
} \\
{\fontsize{11pt}{12pt}\selectfont $^{1}$~State Key Laboratory for Novel Software Technology, Nanjing University, China \quad $^{2}$~Huawei, China} \\
}


\maketitle

\input{sections/abs}

\begin{IEEEkeywords}
Distributed training, fault tolerance, performance modeling, communication optimization. 
\end{IEEEkeywords}

\input{sections/intro}

\input{sections/back}
\input{sections/overview}

\input{sections/design}

\input{sections/eval}

\input{sections/conclusion}

\input{sections/ack}

\bibliographystyle{IEEEtran}
\bibliography{reference}

\end{document}

%% file: sections/abs.tex
\begin{abstract}
Training large language models faces frequent interruptions due to various faults, demanding robust fault-tolerance. Existing backup-free methods—redundant computation, dynamic parallelism, and data rerouting—each incur performance penalties, whether from ongoing overhead, lengthy reconfigurations, or post-recovery inefficiencies. We propose \sysname, an adaptive fault-tolerant system that intelligently selects optimal recovery strategies when a failure occurs.
\sysname achieves this through a unified performance model, quick execution plan search, accurate performance estimation, and efficient communication optimizations. 
Experiments on a 32-card cluster show that \sysname has a performance gap of within 11.00\% between post-recovery and failure-free training, while preserving model convergence and efficient memory usage. 
Compared to state-of-the-art methods, \sysname achieves up to 1.229$\times$ and 1.355$\times$ higher average throughput than Oobleck and Recycle, respectively.
\end{abstract}

%% file: sections/intro.tex
\section{Introduction}
In recent years, large language models have gained significant popularity because of their impressive capabilities in various tasks, such as natural language understanding, code generation, and dialogue systems~\cite{grattafiori2024llama, achiam2023gpt}. 
With the increasing complexity and scale of these models, \emph{long-term and large-scale training} has become a common practice.
For example, training the 405B Llama 3.1 model required roughly 30.8 million GPU hours~\cite{grattafiori2024llama}.
The long-term, large-scale training process can be interrupted on average every few hours due to various faults~\cite{Zhuang2023Gemini}, such as hardware failures, software bugs, or network issues.
Thus, fault-tolerant training techniques are crucial for maintaining model performance and stability, especially in distributed and resource-constrained environments.


In practice, the entire fault-tolerant training process typically consists of three phases. 
(i) \textit{Fault-free phase}: The system trains according to the given configuration when no faults occur. 
(ii) \textit{Fault-handling phase}: When a fault occurs, training is interrupted, and the system takes measures to recover, such as restoring from checkpoints~\cite{Assaf2022Check-N-Run,jiang2024megascalescalinglargelanguage}, backup servers~\cite{Xie2020Elan}, or adjusting the configuration using only the available resources~\cite{thorpe2023bamboo,jang2023oobleck,gandhi2024recycle,wagenlander2024tenplex}. 
(iii) \textit{Post-recovery phase}: After recovery, due to the absence of backups and possible node loss, training efficiency may decrease compared to the fault-free phase. 
Given that many users operate under resource constraints, backup-free fault-tolerant solutions have become a focal point.

However, existing backup-free fault-tolerant training methods are often tailored to specific training phases, leading to inherent limitations in broader scenarios.
First, \textit{redundant computation} such as Bamboo~\cite{thorpe2023bamboo}, where each stage node in pipeline parallelism additionally stores the layers of its successor and performs gradient computation. When the successor fails, its predecessor can take over its tasks seamlessly. However, due to additional computation and memory overhead for each stage, the training throughput in the fault-free phase is reduced.
Second, \textit{dynamic parallelism} like Oobleck~\cite{jang2023oobleck}, which provides predefined pipeline templates. Upon failure, the system can switch to a template with fewer nodes, dynamically adjusting the parallelism to match the available training resources. Although it does not introduce additional overhead during the fault-free phase, it takes a long time to reconfigure the pipelines when a fault occurs, leading to a significant drop in throughput during the fault-handling phase.
Finally, \textit{data rerouting} is exemplified by Recycle~\cite{gandhi2024recycle}, which reroutes the micro-batches assigned to failed nodes to peer nodes in the same pipeline stage but in different data parallel groups, without parameter reconfiguration. However, its training throughput in the post-recovery phase is related to pipeline bubbles, which can lead to inefficiencies if not managed properly. With an increasing number of failures, the degradation of performance becomes more pronounced. 

Therefore, a comprehensive and general fault-tolerant system is urgently needed that is compatible with different training phases.
However, achieving this goal requires overcoming several key challenges.
(i) How to accurately model the performance of different fault-tolerant strategies across all phases to provide a theoretical basis for strategy selection? 
(ii) How to efficiently search for and determine the optimal fault-tolerant strategy and execution plan under various configurations and resource constraints?
and (iii) How to further optimize communication or computation performance for different strategies and training phases to maximize overall training efficiency?

To address these challenges, \sysname introduces a unified framework that integrates performance modeling, strategy selection, and communication optimization for fault-tolerant training. 
Specifically, \sysname builds on a comprehensive performance model that covers all phases and strategies, leveraging both analytical modeling and profiling data to accurately estimate key metrics.
Guided by these estimations, \sysname efficiently searches for the optimal execution plan to maximize overall training throughput under given resource constraints.
In addition, \sysname incorporates targeted optimizations for model transfer during recovery and synchronization communication in data parallelism, further improving the efficiency of the selected strategy. 
With these enhancements, \sysname can achieve efficient and stable fault-tolerant training.

The main contributions of \sysname are as follows.

\head{Performance Modeling}. 
To determine the direction of performance optimization for fault-tolerant training, we first model the performance of different strategies across various training phases, thereby establishing the system's optimization objective. 
In particular, the reconstruction overhead during the fault-handling phase and the training performance after recovery are key factors in our strategy selection. 
We also consider performance modeling in complex scenarios such as multiple failures and asymmetric parallelism, enabling performance evaluation under extreme conditions.
Finally, we provide accurate estimations of per-step execution time and memory usage, offering a solid theoretical foundation for strategy search while effectively avoiding out-of-memory (OOM) issues.

\head{Execution Plan Search}. 
Before selecting a strategy, it is essential to ensure that each candidate strategy is configured for optimal performance. 
For different fault-tolerant strategies, factors such as parallelism degree, failed node distribution, model partitioning, and data allocation can significantly affect post-recovery training efficiency. 
To address this, \sysname employs a heuristic search that comprehensively considers these factors while controlling search overhead, enabling rapid identification of feasible execution plans. 
\sysname selects the optimal plan based on estimated performance metrics.

\head{Communication Optimization}. 
Changes of the execution plan not only affect pipeline computation but also inevitably impact communication. 
These impacts fall into two categories: (i) weight transfer communication during training reconstruction in dynamic parallelism; 
(ii) synchronization communication in data parallelism, which may be affected by asymmetric parallel execution. 
Both types of communication overhead are dynamic and, if not handled properly, can incur high costs. 
Therefore, we optimize both types of communication: the former is abstracted as a bipartite graph matching problem, and the latter as a graph coloring problem, achieving optimal communication performance with minimal overhead.

\sysname was extensively evaluated in both simulated and real-world environments, utilizing a cluster of 32 Ascend 910B AI training accelerators (NPUs)~\cite{liao2021ascend}. 
Compared to fault-free training, \sysname's post-recovery performance maintained a gap of less than 11.00\%. 
In simulations, \sysname consistently outperformed baselines like Oobleck and Recycle, achieving an average throughput of 1.229$\times$ and 1.355$\times$ higher, respectively. 
A detailed evaluation of \sysname's key techniques shows that its estimator achieves an accuracy error within 8.02\%. 
Optimized weight transfer reduces recovery time by up to 26.79\%, while asymmetric communication optimization shortens step time by up to 15.44\%. 
Memory, convergence, and scalability analyses show that \sysname performs well in all three aspects.

%% file: sections/back.tex
\section{Background}
In this section, we introduce the necessary background of distributed model training, state-of-the-art backup-free work for fault tolerance, and the limitations of these methods. 

\subsection{Distributed Model Training}
With the rapid growth of model sizes, distributed training leverages strategies including Data (DP)~\cite{goyal2018accuratelargeminibatchsgd,sergeev2018horovodfasteasydistributed,you2018imagenettrainingminutes}, Pipeline (PP)~\cite{smith2022usingdeepspeedmegatrontrain,huang2019gpipeefficienttraininggiant,kim2023memorybalancedpipelineparallelism}, Tensor (TP)~\cite{rajbhandari2020zeromemoryoptimizationstraining}, and Expert Parallelism (EP)~\cite{cai2025shortcutconnectedexpertparallelismaccelerating,qian2025epsmoeexpertpipelinescheduler,liu2025moeparallelfoldingheterogeneous}.
\textit{DP} involves splitting the dataset across multiple nodes, where each node processes a subset of the data and computes the gradients independently.
\textit{PP} divides the model into stages, with each stage running on a different node, while \textit{TP} splits the model's tensors across multiple nodes, allowing for parallel computation of large tensors.
And \textit{EP}, often used in the mixture of experts (MoE) models~\cite{Fedus2022Switchtransformers}, assigns different experts across nodes, enabling the model to handle more complex tasks by activating only a subset of experts.
These parallel strategies have different communication patterns, which determine their typical deployment scenarios. 
Specifically, \textit{TP} and \textit{EP} require frequent and fine-grained communication, making them more suitable for intra-node deployment~\cite{jin2025megascalemoelargescalecommunicationefficienttraining}.  
In contrast, \textit{DP} and \textit{PP} are more suitable for inter-node scaling due to their lower communication overhead~\cite{Li2024UnderstandingCommunicationCharacteristicsofDistributedTraining,Moolchandani2023AMPeD}.
In practice, these strategies can be organized properly to achieve hybrid parallelism~\cite{korthikanti2022reducingactivationrecomputationlarge,Li2021Chimera}.
While TP and EP are crucial for scaling model capacity and throughput, their fine-grained communication and tight cross-device coupling make fault isolation and recovery particularly challenging. Consequently, most fault-tolerant systems~\cite{athlur2022varuna,thorpe2023bamboo,jang2023oobleck,gandhi2024recycle,kang2025elaswaveelasticnativescalablehybridparallel} focus on DP and PP, utilizing their natural state partitioning for flexible fault isolation with minimal disruption.

\subsection{Fault Tolerance in Distributed Training}
Long-running large-scale distributed training jobs are susceptible to various faults~\cite{Gupta2017Failuresinlargescalesystems,huang2019gpipeefficienttraininggiant,weng2022MLaaSintheWild}, such as hardware failures, network issues, and software bugs. 
For example, during the 54-day Llama 3.1 training~\cite{grattafiori2024llama}, there were 419 unplanned interruptions caused by faults, averaging 3 hours between each interruption.
Gemini~\cite{Zhuang2023Gemini} noted that during the OPT-175B model training on 992 A100 GPUs, approximately 110 faults occurred over two months. 
Minder~\cite{Deng2025Minder} showed that when training tasks involved about 1,000 machines, approximately 2 faults occurred daily on average, with each fault potentially halting training for several hours. 
To mitigate the impact of faults, various fault-tolerance strategies have been proposed. 
A common approach is \textit{checkpointing}~\cite{Assaf2022Check-N-Run,jiang2024megascalescalinglargelanguage}, which periodically saves the model state to disk, allowing recovery from the last checkpoint in case of interruptions.
Another strategy to ensure rapid recovery is \textit{warm backup}~\cite{Xie2020Elan}, which maintains a pool of idle nodes ready to take over tasks at any time. However, it incurs significant resource waste due to the low utilization of standby nodes. 
As a result, this paper focuses on the \textit{backup-free} methods, which leverage inherent redundancy and dynamic reconfiguration within the active training cluster and are orthogonal to checkpoint- and backup-based solutions.

\subsection{Limitations of Backup-free Methods}
In practice, state-of-the-art backup-free methods can be categorized into three main categories:
\begin{itemize}
\item \textbf{Redundant computation}. 
Bamboo~\cite{thorpe2023bamboo} employs redundant computation within the training pipeline for fault tolerance. 
Each node maintains a replica of its successor and hides the computation overhead within the pipeline bubbles, enabling rapid recovery for a node failure.

\item \textbf{Dynamic parallelism} means adjusting the parallelism to adapt to the available resources.
Varuna~\cite{athlur2022varuna} proposed adjusting the pipeline and data parallelism to minimize communication overhead and recover by continuously checkpointing the model states.
Oobleck~\cite{jang2023oobleck} introduced pipeline templates to run multiple replicas of the pipeline and reconstruct the lost stage from the redundant replicas.

\item \textbf{Data rerouting}. 
Recycle~\cite{gandhi2024recycle} exploits the consistent parameters across different pipelines and reroutes the micro-batches on the failed node to its data-parallel peers.
Further, it proposed a scheduling mechanism to allow rerouted micro-batches to execute within the inherent pipeline bubbles, minimizing the throughput degradation.
\end{itemize}

We analyze the performance impact of these backup-free methods throughout the training process, which includes three critical phases:
(i) \textit{Fault-free phase}. A fault-tolerance policy may also introduce additional overhead when no faults occur.
(ii) \textit{Fault-handling phase}. This phase typically includes searching for a new execution plan and restoring the training state.
(iii) \textit{Post-recovery phase}. After recovering from the failure, the model's training efficiency may be degraded. 
The time and memory impact of different fault-tolerance policies varies.
Redundant computation minimizes fault-handling time but introduces a prohibitive overhead (nearly double the computation and memory) during the fault-free phase.
Dynamic parallelism methods avoid overhead during fault-free training. The primary drawback is the high reconstruction overhead during fault-handling, such as restarting from checkpoints or copying from live replicas, especially with frequent failures.
Data rerouting avoids huge reconstruction overhead by rerouting micro-batches on the fly. However, its effectiveness relies on complex scheduling to hide additional computation within pipeline bubbles. If the bubbles are highly optimized (e.g., zero-bubble parallelism~\cite{qi2023zerobubble}), the performance loss after resuming training is not negligible.
In summary, existing methods often only consider performance in certain phases and lack a comprehensive solution.

The challenge hinges on balancing immediate fault-handling overhead with post-recovery performance. 
Data rerouting is preferable for individual failures due to its low handling cost. However, as failures accumulate, training performance after recovery can become unsatisfactory. 
Conversely, while dynamic parallelism has a higher reconstruction cost, it may achieve better long-term throughput post-recovery by a new optimal parallel plan. Therefore, we consider selecting the optimal fault-tolerance strategy in real time based on specific failure scenarios to ensure efficient and stable training.

%% file: sections/overview.tex
\section{Performance Modeling}\label{sec:overview}
In this section, we first model the performance of targeted fault-tolerance methods, including data rerouting and dynamic parallelism, and then determine the optimization goal. 
Based on these, we present our system \sysname, which adaptively selects the most suitable fault-tolerance policy.


\begin{Definition}[State]
    The state of a training system, denoted as $\mathcal{S}$ includes the following information:
    \begin{itemize}
        \item \textbf{Cluster Status}: The number of nodes in the cluster. 
        \item \textbf{Execution Plan}:
        (i) the fault-tolerance policy, which can be either dynamic parallelism or data rerouting;
        (ii) the parallel configuration, including the number of data parallel groups $N_{dp}$ and the number of pipeline stages $N_{pp}$;
        (iii) the data distribution among different data parallel groups, where the sum equals the global batch size;
        (iv) the model layer distribution across pipeline stages;
        (v) the distribution of failed nodes across the stages.
        Detailed information is provided in Section~\ref{sec:planner}.
    \end{itemize}
\end{Definition}



Therefore, when a fault occurs, the system needs to switch from the current state $\mathcal{S}_1$ to a new state $\mathcal{S}_2$. 
Generally, considering a series of faults or recovery operations, we can define the state transition as a sequence of states $\mathcal{S}_1, \mathcal{S}_2, \ldots, \mathcal{S}_n$.

Considering the whole training process, the number of processed samples $S_{\text{total}}$ can be calculated as follows:
\begin{equation}
    S_{\text{total}} = \sum_{i=1}^{n} S_i ,
\end{equation}
where $S_i$ is the number of samples processed in state $\mathcal{S}_i$ and $t_{\mathcal{S}_i}$ is the time spent in state $\mathcal{S}_i$.
Moreover, the execution time $t_{\text{total}}$ of the whole training process can be calculated as:
\begin{equation}
    t_{\text{total}} = \sum_{i=1}^{n} t_{\mathcal{S}_i} + \sum_{i=1}^{n-1} t_{\mathcal{S}_i \to \mathcal{S}_{i+1}},
\end{equation}
where $t_{\mathcal{S}_i \to \mathcal{S}_{i+1}}$ is the time spent in transitioning from state $\mathcal{S}_i$ to $\mathcal{S}_{i+1}$.

\head{Optimization Problem.} 
Straightforwardly, we aim to maximize the throughput of the whole training process, which can be formulated as:
\begin{equation}
    \mathop{\arg\max}  _{\mathcal{S}_1, \mathcal{S}_2, \ldots, \mathcal{S}_n} \frac{S_{\text{total}}}{t_{\text{total}}} 
\end{equation}
However, it is impractical to optimize the whole training process, as it requires knowledge of the entire training process, including when the next fault will occur.
Fortunately, we observe that \emph{the duration between two faults, i.e., $t_{\mathcal{S}_i \to \mathcal{S}_{i+1}}+t_{\mathcal{S}_{i+1}}$, remains constant\footnote{Note that for different durations, e.g., $j\neq i$, the duration can be different, but the duration between two given faults is constant.} regardless of the execution plan, and changing the execution plan only affects the throughput of the current state $\mathcal{S}_i$ and the transition time $t_{\mathcal{S}_i \to \mathcal{S}_{i+1}}$}.

Therefore, the optimization goal can be simplified to maximizing the throughput of each duration between two faults, which can be expressed as:
\begin{equation}
    \forall i \in [1, n-1], \quad \mathop{\arg\max}_{\mathcal{S}_i} \frac{S_{i+1}}{t_{\mathcal{S}_i\to \mathcal{S}_{i+1}}+ t_{\mathcal{S}_{i+1}}}.
\end{equation}
Note that for the first state $\mathcal{S}_1$, its throughput can be determined by the initial configuration, which has various work to discuss, such as Alpa~\cite{Zheng2022Alpa}, and we omit the discussion here.

For the number of processed samples $S_i$, it can be calculated as follows:
\begin{equation}
    S_i = T_i \cdot t_{\mathcal{S}_i},
\end{equation}
where $T_i$ is the throughput of state $\mathcal{S}_i$.

By substituting $S_i$ into the optimization goal, we can obtain:
\begin{equation}
    \forall i \in [1, n-1], \quad \mathop{\arg\max}_{\mathcal{S}_{i+1}} \frac{T_{i+1} \cdot t_{\mathcal{S}_{i+1}}}{t_{\mathcal{S}_i\to \mathcal{S}_{i+1}} + t_{\mathcal{S}_{i+1}}}.
\end{equation}

Considering the $T_{i}$ in state $\mathcal{S}_i$ is calculated as:
\begin{equation}
    T_i = \frac{B_i}{t_{\text{step}, \mathcal{S}_i}},
\end{equation}
where $B_i$ is the given global batch size in state $\mathcal{S}_i$, and $t_{\text{step}, \mathcal{S}_i}$ is the time for each step in state $\mathcal{S}_i$.

Therefore, we can further express the optimization goal as:
\begin{equation}
    \forall i \in [1, n-1], \quad \mathop{\arg\max}_{\mathcal{S}_{i+1}} \underbrace{\frac{B_{i+1}}{t_{\text{step}, \mathcal{S}_{i+1}}}}_{\text{throughput}} \cdot \underbrace{\frac{t_{\mathcal{S}_{i+1}}}{t_{\mathcal{S}_i\to \mathcal{S}_{i+1}} + t_{\mathcal{S}_{i+1}}}}_{\text{effective time ratio}},
\end{equation}
which mainly consists of two parts: the throughput of the next state $\mathcal{S}_{i+1}$ and the effective time ratio, which represents the proportion of time spent in state $\mathcal{S}_{i+1}$ relative to the total time between two faults. Regarding the throughput, since the batch size $B_{i+1}$ is given by users, the key to maximizing the throughput is to minimize the step time $t_{\text{step}, \mathcal{S}_{i+1}}$, which represents the time for each training step (a.k.a., iteration).
For the effective time ratio, since the total time between two faults is $t_{\mathcal{S}_i \to \mathcal{S}_{i+1}} + t_{\mathcal{S}_{i+1}}$, which is constant, the key to maximizing the effective time ratio is to minimize the state transition time $t_{\mathcal{S}_i \to \mathcal{S}_{i+1}}$. 
We need to investigate the factors that affect the state transition time $t_{\mathcal{S}_i \to \mathcal{S}_{i+1}}$ and the per-step execution time $t_{\text{step}, \mathcal{S}_{i+1}}$. 
First, the fault-tolerance strategy in the execution plan allows for a choice between data rerouting and dynamic parallelism. 
The transition time $t_{\mathcal{S}_i \to \mathcal{S}_{i+1}}$ varies significantly between these strategies. 
For example, data rerouting does not require training reconstruction, so the transition time is negligible. 
In contrast, dynamic parallelism incurs considerable transition time due to weight transfer between nodes and restart overhead. 
The factors affecting $t_{\text{step}, \mathcal{S}_{i+1}}$ are more complex, involving parallel configuration, as well as distribution of failed nodes, data, and model layers.

In summary, we can abstract the problem of fault-tolerant training as follows: Given the current state $\mathcal{S}_i$, how can we determine the next state $\mathcal{S}_{i+1}$ (specifically, its execution plan $E_i$) such that both the state transition time $t_{\mathcal{S}_i \to \mathcal{S}_{i+1}}$ and the per-step execution time $t_{\text{step}, \mathcal{S}_{i+1}}$ are minimized? 
To address this, we design the \sysname system, which adaptively finds the most suitable execution plan in response to failures. 
Specifically, \sysname tackles the following challenges.

\begin{itemize}
    \item \textbf{Given $\mathcal{S}_i$, how do we determine the optimal  $\mathcal{S}_{i+1}$? (\S\ref{sec:planner})} We employ a heuristic search that rapidly explores candidate execution plans considering multiple factors, including parallelism, data, and layer distribution. 
    We select the optimal execution plan by evaluating the estimated $t_{\text{step}, \mathcal{S}_{i+1}}$ and $t_{\mathcal{S}_i \to \mathcal{S}_{i+1}}$ of each plan.

    \item \textbf{How do we minimize the state transition time $t_{\mathcal{S}_i\to\mathcal{S}_{i+1}}$? (\S\ref{sec:restorer})} During dynamic parallelism, $t_{\mathcal{S}_i \to \mathcal{S}_{i+1}}$ includes the weight transfer overhead incurred by training reconstruction, which varies with the execution plan. 
    We abstract this process as a bipartite matching problem to minimize transfer cost, thus reducing $t_{\mathcal{S}_i \to \mathcal{S}_{i+1}}$.
    
    \item \textbf{How do we minimize the step time after recovery $t_{\text{step}, \mathcal{S}_{i+1}}$? (\S\ref{sec:restorer})} The $t_{\text{step}, \mathcal{S}_{i+1}}$ consists of both pipeline computation time and synchronization communication time. 
    While the former can be estimated by the estimator, the latter, especially under asymmetric parallelism, can be optimized as a graph coloring problem.
    
    \item \textbf{How do we estimate the step time after recovery $t_{\text{step}, \mathcal{S}_{i+1}}$? (\S\ref{sec:estimator})} \sysname estimates both step time and memory usage for each candidate plan. 
    For the former, it predicts pipeline time by applying analytical formulas for symmetric pipelines (data rerouting) and dynamic programming for asymmetric pipelines (dynamic parallelism). 
    For the latter, it identifies the peak memory of each pipeline stage based on layer-wise profiling, ensuring that no OOMs occur.
\end{itemize}

%% file: sections/design.tex
\section{System Design}\label{sec:design}

\begin{figure}
    \centering
    \includegraphics[width=0.75\linewidth]{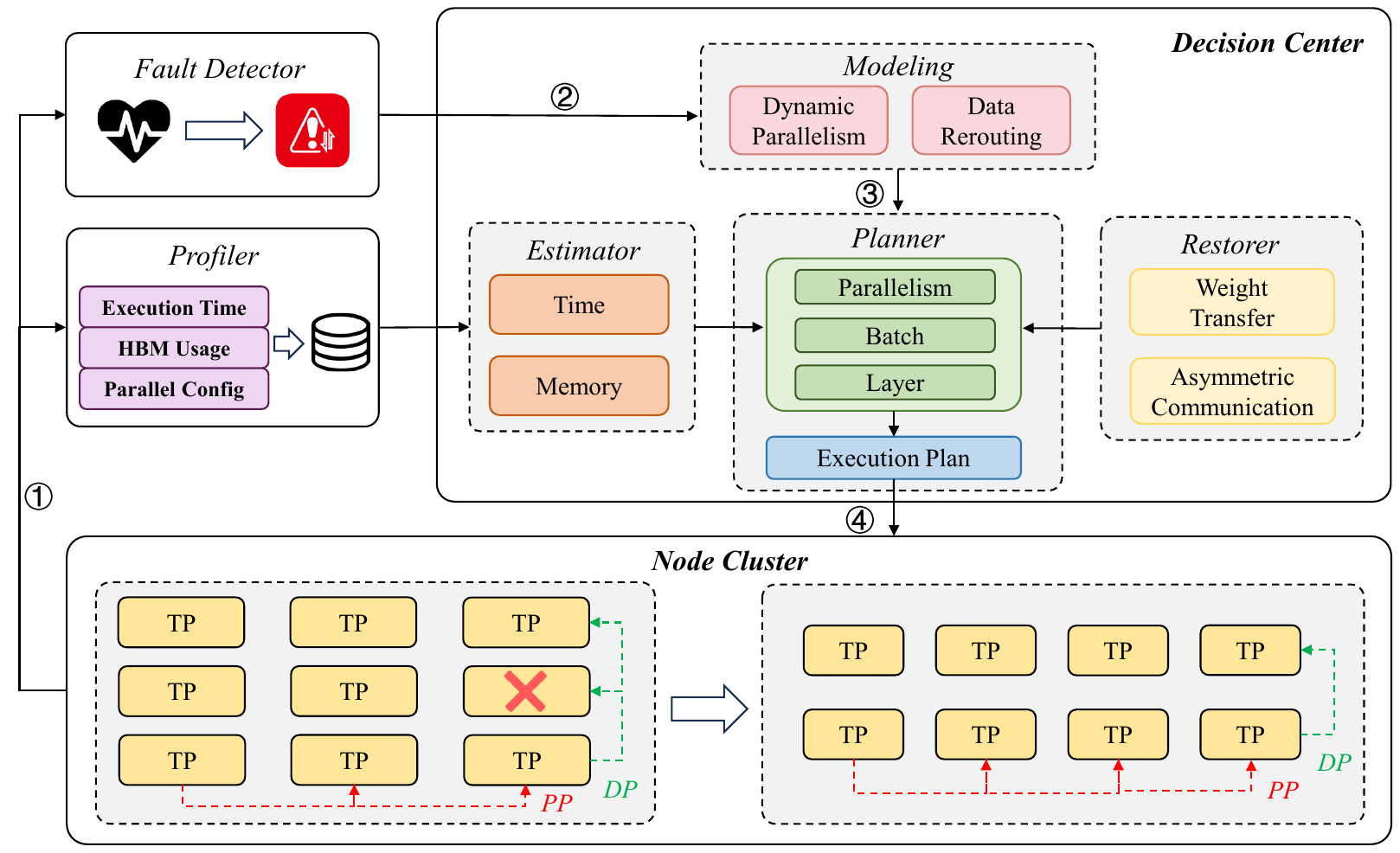}
    \caption{The overall workflow of \sysname.}
    \vspace{-0.2in}
    \label{fig:workflow}
\end{figure}

In this section, we will introduce the core techniques of \sysname and explain how they enable fast recovery and efficient training.
As shown in Figure~\ref{fig:workflow}, \sysname consists of three main components: a profiler, a fault detector, and a decision center. The decision center integrates a planner, estimator, and restorer to manage the recovery process. The workflow is as follows:
\textcircled{1} \textit{Monitoring}: The profiler continuously collects runtime metrics such as execution time, HBM usage, and parallel configurations from the cluster. Meanwhile, the fault detector monitors the health of each node.
\textcircled{2} \textit{Fault Trigger}: Upon detecting a failure, the fault detector immediately triggers the decision center to initiate the fault-tolerance mechanism.
\textcircled{3} \textit{Decision Making}: For each failure, the planner generates potential execution plans considering parallelism, layer, and batch configurations. Then it queries the estimator, which uses profiling data to predict the performance and memory usage of each plan.
Meanwhile, the restorer is responsible for minimizing the overhead of transferring model weights and synchronization communication in asymmetric parallelism. 
Based on these, the planner selects the optimal execution plan.
\textcircled{4} \textit{Plan Execution}: The planner sends the finalized execution plan to the cluster, which isolates the failed node and resumes training with minimal disruption.



\subsection{Planner}\label{sec:planner}
In the planner, we need to determine the execution plan that achieves the best performance.

\begin{figure}
    \centering
    \includegraphics[width=0.75\linewidth]{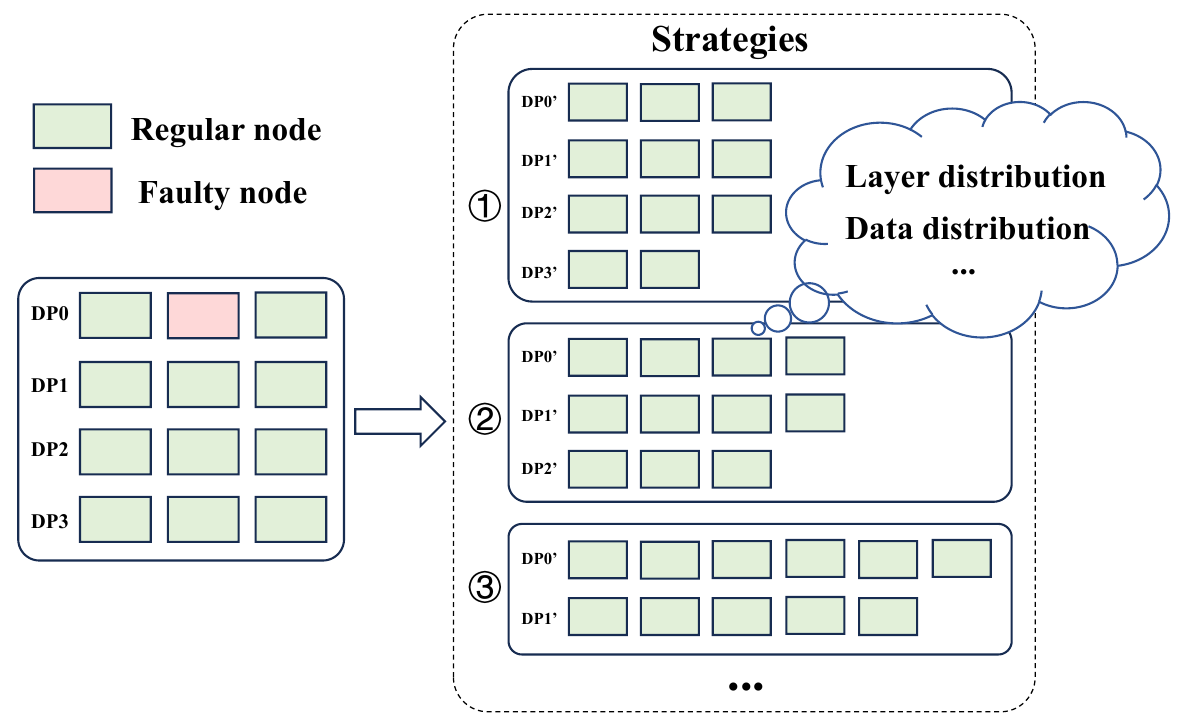}
    \caption{Search for optimal execution plan.}
    \vspace{-0.2in}
    \label{fig:search}
\end{figure}

\head{Execution Plan Search}.
If data rerouting is selected, considering that pipeline bubbles during training are usually well-optimized and the proportion is limited, the upper and lower bounds of performance do not differ significantly. 
Therefore, when determining the execution plan, we evenly distribute the micro-batches from the failed nodes to other data-parallel peers, aiming to overlap them with bubbles as much as possible. 
However, if dynamic parallelism is used, the performance differences between different execution plans can be substantial. As shown in Figure~\ref{fig:search}, we need to search for the optimal execution plan based on the following factors.

\begin{itemize}
    \item \textbf{DP/PP Parallelism}. Unlike directly reducing one DP or PP degree, our goal is to utilize all available nodes as much as possible to maximize throughput. 
    Thus, we provide a larger parallelism search space that supports asymmetric parallelism and is not restricted by predefined PP templates as in Oobleck. 
    Based on our experience, the new DP degree often differs from the original by less than 2. We employ a heuristic search by limiting the parallelism range to reduce the search overhead.       
    
    \item \textbf{Batch Distribution}. For load balancing of different DP groups and to avoid stragglers, we first pre-allocate micro-batches according to the proportion of nodes. 
    If there are remaining unallocated micro-batches, we recursively assign them one by one to all pipelines.
    If some partitions have no data finally, we reassign one micro-batch from the largest partition to fill these. This check is repeated to ensure that no partition is left idle.

    \item \textbf{Layer Distribution}. For a given pipeline, special attention must be paid to load balancing between stages. Therefore, we first evenly split the model layers across all stages. For the remaining layers that cannot be evenly divided, we enumerate all possible allocation schemes. 
    After filtering out options that may cause OOM via memory estimation, we select the splitting scheme with the lowest pipeline execution time by the estimator.
\end{itemize}

Algorithm~\ref{alg:search} illustrates the steps to identify the optimal execution plan, with $get\_execution\_plan$ serving as entry.
Lines 1-7 demonstrate the function $get\_parallel\_strategy$ to enumerate all candidate parallel strategies ($S_{cand}$), taking into account possible failure numbers and valid DP/PP combinations. Specifically, it recursively factorizes the current number of nodes to determine the possible degrees of DP/PP that meet the required range in $integer\_partition$.
In the $get\_execution\_plan$ function, for each candidate strategy in $S_{cand}$, lines 13-14 determine the balanced distribution of micro-batches ($distribute\_batch$) and the optimal splitting of layers into stages ($split\_layers$).
Line 15 estimates the execution time of the candidate by $time\_estimator$.
The function then tracks the plan ($P_{best}$) with the lowest estimated execution time and returns it in lines 16-19.

\begin{algorithm}[htbp]
    \caption{Search for the Best Execution Plan}
    \label{alg:search}
    \DontPrintSemicolon
    
    \KwIn{
        Number of Nodes $N$, 
        Max Number of Faults $N_f$, 
        Range of DP $R_{dp}$, 
        Range of PP $R_{pp}$,
        Micro Batches $N_m$
    }
    \KwOut{The Best Execution Plan $P_{best}$}

    \SetKwProg{Fn}{Function}{}{end} 
    \Fn{get\_parallel\_strategy($N, N_f, R_{dp}, R_{pp}$)}{
        $S_{cand} = \text{null}$\;
        \For{$i \leftarrow 1$ \KwTo $N_f$}{
            \For{each $dp$ in $R_{dp}$}{
                $L_{pp} = \text{integer\_partition}(N-i, dp, R_{pp})$\;
                $S_{cand} \mathbin{+\!=} (dp, L_{pp})$\;
            }
        }
        \KwRet{$S_{cand}$}\;
    }
    
    \Fn{get\_execution\_plan($N, N_f, R_{dp}, R_{pp}, N_m$)}{
        $T\_{best} = \infty$\;
        $P\_{best} = \text{null}$\;
        $S_{cand} = \text{get\_parallel\_strategy}(N, N_f, R_{dp}, R_{pp})$\;
        \For{each $(dp, L_{pp})$ in $S_{cand}$}{
            $batch = \text{distribute\_batch}(N_m, L_{pp})$\;
            $layers = \text{split\_layers}(L_{pp}, \text{memory\_estimator})$\;
            $T_{est} = \text{time\_estimator}(L_{pp}, layers, batch)$\;
            \If{$T_{est} < T_{best}$}{
                $T_{best} = T_{est}$\;
                $P_{best} = (dp, L_{pp}, layers, batch)$\;
            }
        } 
        \KwRet{$P_{best}$}\;
    }
\end{algorithm}
\vspace{-0.1in}

The Planner's overhead involves three steps: (i) Parallel strategy search iterates through valid $(dp, L_{pp})$ combinations based on the divisors of the node count $N$. The complexity is $O(\tau(N))$, where $\tau(n)$ denotes the divisor function. Even for large clusters (e.g., $N=1024$), the number of divisors is small (e.g., $\tau(1024)=11$). (ii) The complexity of batch distribution is linear with respect to the number of data-parallel groups, $O(N_{dp})$. (iii) Due to the even-distribution-first policy, layer distribution is limited to allocating the remaining layers $r = L \mod N_{pp}$ across stages. The complexity is bounded by the binomial coefficient $\binom{N_{pp}}{r}$. Since the pipeline depth $N_{pp}$ is typically small (e.g., $\le 16$) in practice, the number of candidates remains in the range of ten to hundreds. Furthermore, \sysname pre-calculates and caches the optimal plans for potential failure scenarios (e.g., loss of 1 to $k$ nodes) during the fault-free training phase. When a fault occurs, the Restorer directly retrieves the pre-computed plan from the cache, making the decision-making latency negligible. 

\subsection{Restorer}\label{sec:restorer}
There is significant room for optimization of $t_{\mathcal{S}_i\to\mathcal{S}_{i+1}}$ and $t_{\text{step}, \mathcal{S}_{i+1}}$. 
For the former, it suffers from the communication overhead brought by weight transfer, while the latter may suffer from asymmetric synchronous communication.

\begin{figure}[tbp]
    \centering
    \includegraphics[width=0.8\linewidth]{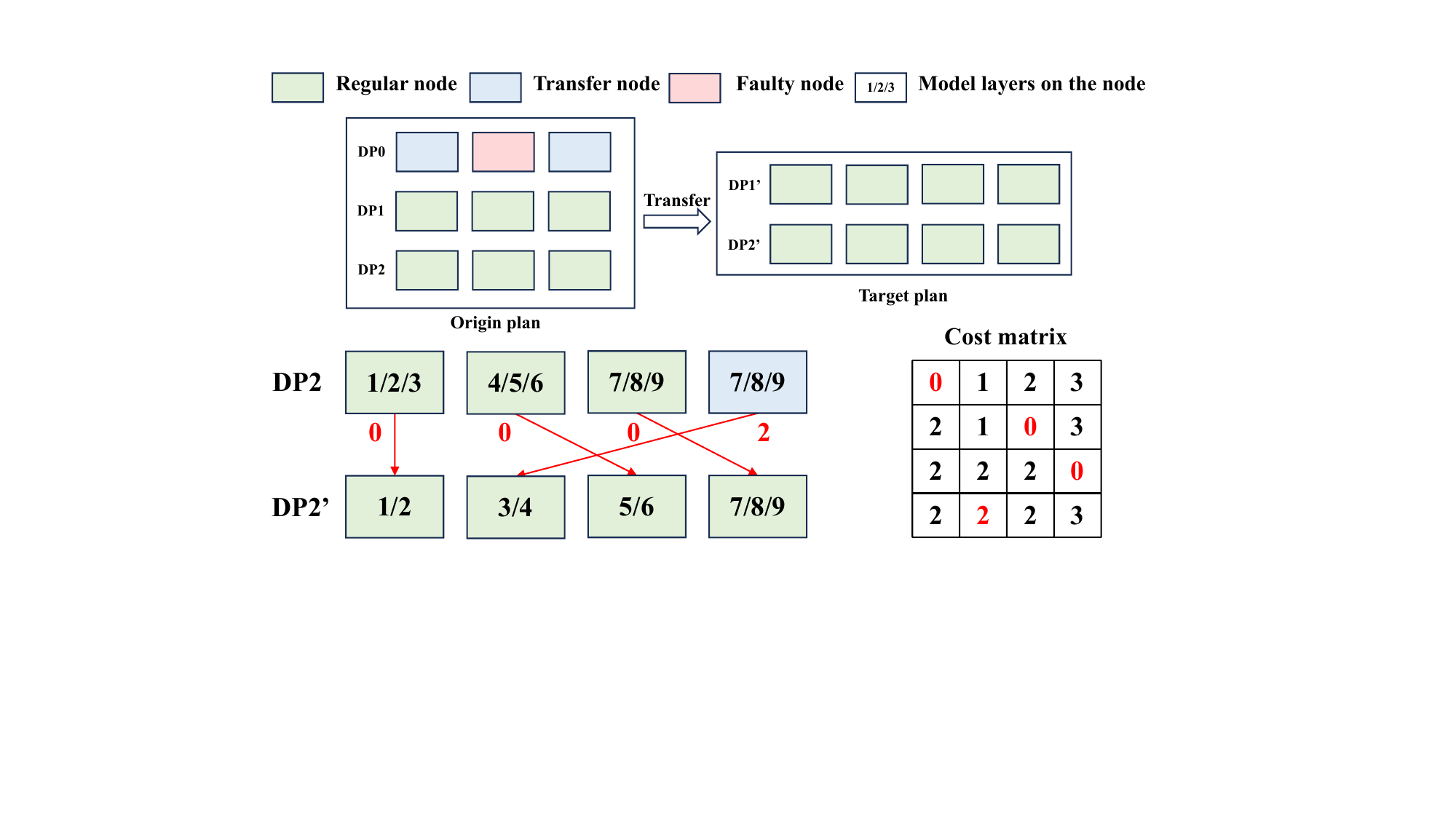}
    \caption{Optimization of weight transfer.}
    \label{fig:weight_transfer}
    \vspace{-0.2in}
\end{figure}

\head{Weight Transfer}. To achieve higher training throughput, we prefer to use dynamic parallelism to rearrange the training nodes. 
However, the high overhead of model weight transmission can become a bottleneck in practical deployment. 
As shown in Figure~\ref{fig:weight_transfer}, the original training uses a configuration of DP = 3 and PP = 3, with 9 layers of model weights evenly distributed in each stage. When a node failure occurs, the planner identifies an execution plan, i.e., DP = 2 and PP = 4, with the model layers distributed as (2,2,2,3). 
To migrate the layers from the original to the new configuration, how can we minimize the amount of data transferred? We can model this as a bipartite graph matching problem.

Taking the original $DP2$ group and the new $DP2'$ group as an example, there may be different correspondences between the two, resulting in varying amounts of weight data that need to be transferred. 
Assuming the number of remaining nodes is $N$, we can construct an $N \times N$ cost matrix $Cost$ based on the different layer distributions, where $Cost[i][j]$ represents the cost for node $i$ to migrate to the $j$-th node under the new plan.
For example, if the first node in $DP2$ corresponds to the first node in $DP2'$, the layers change from (1,2,3) to (1,2); layer 3 can be directly discarded without any additional data transfer, so the cost is 0. 
However, if this node corresponds to the second node in $DP2'$, the layers become (3,4); in addition to discarding layers 1 and 2, layer 4 must be transferred from another node, resulting in a cost of 1.
In this way, we can further use the Kuhn-Munkres algorithm~\cite{Kuhn2012Kuhn–Munkres} to compute the migration scheme with the minimum total cost.

\begin{figure}[tbp]
    \centering
    \includegraphics[width=0.7\linewidth]{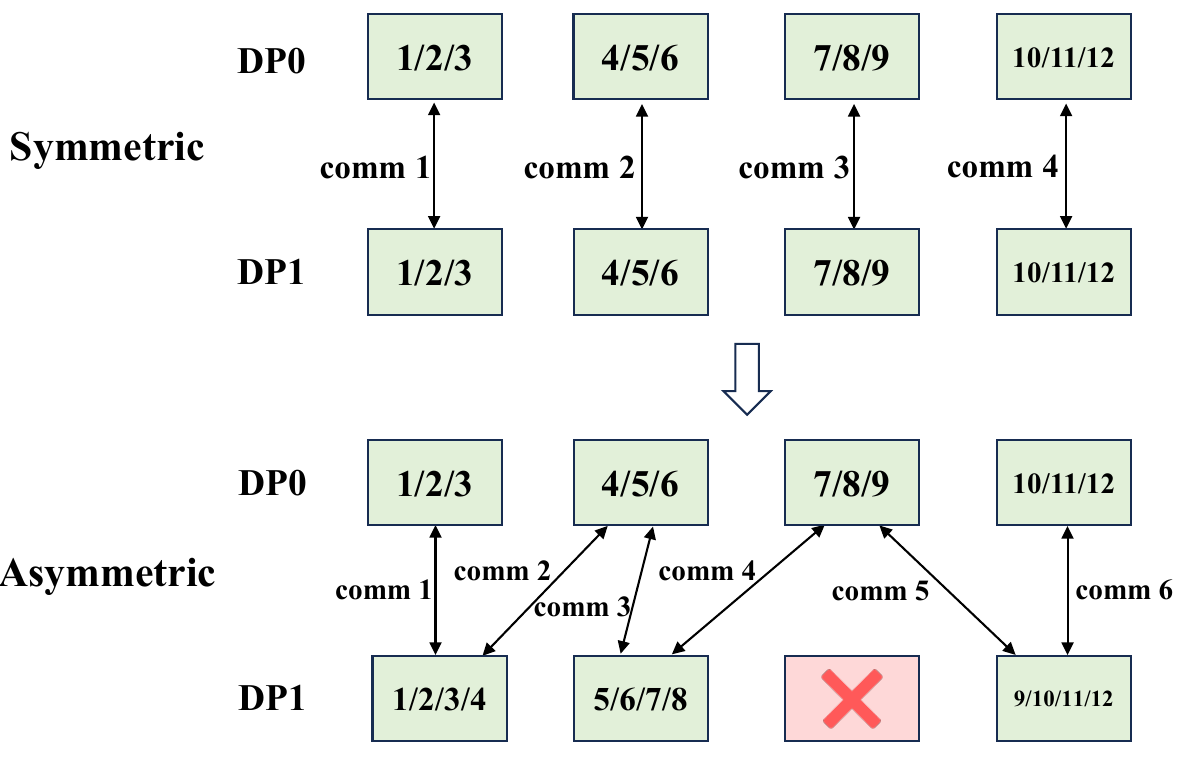}
    \caption{The asymmetric DP gradient update communication.}
    \label{fig:asymmetric_communication}
    \vspace{-0.2in}
\end{figure}

\head{Asymmetric Communication}. During dynamic parallelism, asymmetric parallelism may introduce additional communication overhead for $t_{\text{step}, \mathcal{S}_{i+1}}$. As shown in Figure~\ref{fig:asymmetric_communication}, in symmetric parallel training, AllReduce communication between DP groups occurs for the same model layer. However, when a node failure causes the PP configuration in one DP group to change, the model layers are redistributed. 
To ensure communication between the same layers, we need to establish asymmetric communication domains, which mainly leads to two changes as shown in Figure~\ref{fig:asymmetric_communication}: (i) the number of communications increases (e.g., from 4 AllReduce operations to 6); (ii) there are dependencies between multiple DP communications, causing originally parallel communications to be executed serially. For example, in the new comm2, layer 4 is distributed on the first node, so comm2 must wait for comm1 on that node to finish before it can proceed.

To minimize the total time for asymmetric communication, the key is to maximize the parallelism of DP communications that do not have dependencies. 
We model this problem as a graph coloring problem: each model layer is treated as a vertex, and if two layers are located on the same device, there is a communication dependency, represented by an edge. 
The goal is to assign different colors to adjacent vertices, using the minimum number of colors. Layers with the same color can perform DP communication in parallel, and the number of colors corresponds to the number of communication rounds required. 
We adopt a greedy algorithm to find the minimum number of communication rounds, with a time complexity of $O(L^2)$, where $L$ is the number of model layers.

Besides model weights, our Restorer also incorporates the transfer and reconstruction of optimizer states. For ZeRO parallelism, we only focus on scenarios where state redundancy exists; in such cases, lost optimizer states are recovered by directly transferring replicas from peer nodes in other healthy DP groups, thereby ensuring lossless precision.

\subsection{Estimator}\label{sec:estimator}
During the selection of fault-tolerance policy, especially given a specific execution plan, we need to estimate its performance in terms of time and memory overhead.
In practice, its time should be minimized as much as possible, while its memory usage needs to stay within the hardware limit. 

\head{Time Estimator}. 
\textit{Dynamic parallelism}. When faced with failures, the system can adjust the data and pipeline parallelism sizes to better utilize available resources.
Two scenarios exist:
(i) \textit{Symmetric parallelism}: In Varuna and Parcae, the new parallelism configuration requires the number of nodes to be a multiple of the new DP and PP sizes, ensuring an even distribution of data and model across nodes to maximize throughput.
(ii) \textit{Asymmetric parallelism}: Oobleck provides predefined pipeline templates, each specifying a different number of pipeline stages. In certain parallel configurations, multiple templates may coexist, resulting in asymmetric parallelism. For an irregular number of nodes, this approach can effectively utilize all available nodes.

\begin{figure}[ tbp]
    \centering
    \begin{subfigure}{0.8\linewidth}
        \centering
        \includegraphics[width=\linewidth]{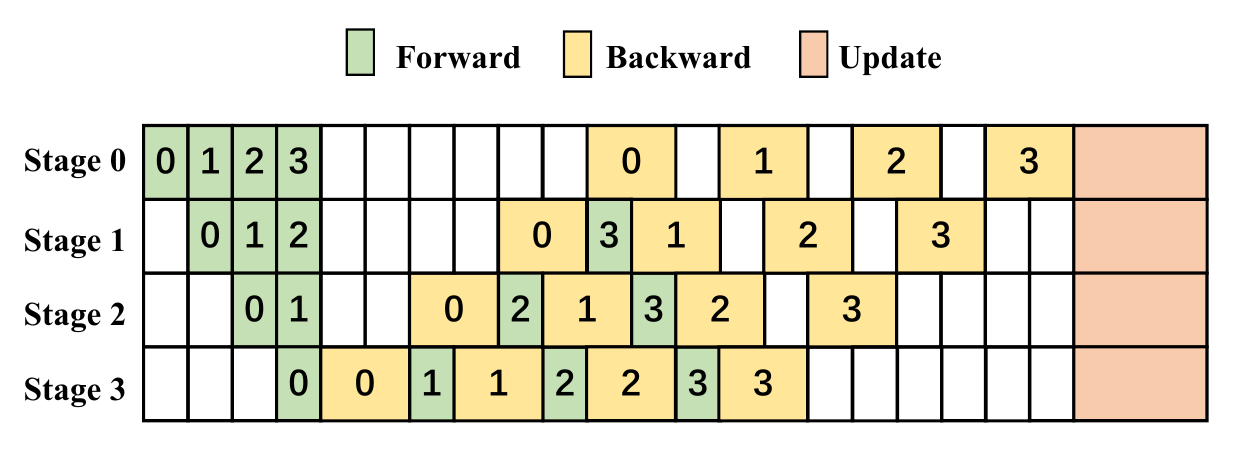}
        \caption{Execution of symmetric pipeline.}
        \label{fig:pipeline}
    \end{subfigure}
    \vspace{0.5em}
    \begin{subfigure}{0.8\linewidth}
        \centering
        \includegraphics[width=\linewidth]{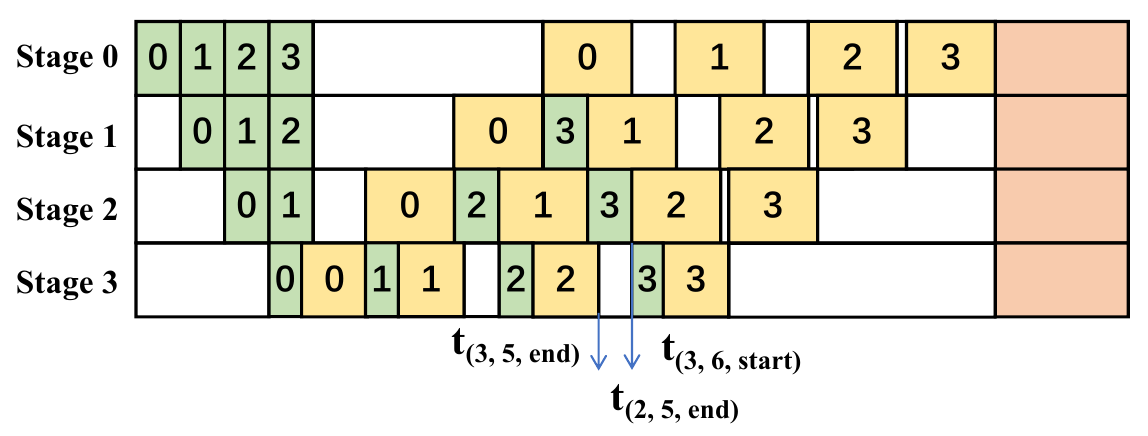}
        \caption{Execution of asymmetric pipeline.}
        \vspace{-0.1in}
        \label{fig:complex_pipeline}
    \end{subfigure}
    \caption{Execution of different pipeline scenarios.}
    \vspace{-0.2in}
    \label{fig:pipeline_combined}
\end{figure}

Different parallelism strategies yield different $t_{\text{step}, \mathcal{S}_{i+1}}$. 
Take the 1F1B pipeline parallelism as an example, as shown in Figure~\ref{fig:pipeline}, the training consists of $N_{dp}$ pipelines, each pipelined into $N_{pp}$ stages. In each iteration, every stage processes $N_m$ micro-batches, with forward and backward pass times per micro-batch denoted as $T_f$ and $T_b$, respectively.
The execution of pipelines shows different patterns under symmetric and asymmetric parallelism.
For the former, the step time can be calculated as follows:
\begin{equation}
t_{\text{step}, \mathcal{S}_{i+1}} = (N_{pp} + N_m - 1) \cdot (T_f + T_b) \label{eq:original_step_time}
\end{equation} 
However, for the latter, the step time is determined by the slowest pipeline due to the synchronous update across all pipelines. 
This relationship is formulated as:
\begin{align}
t_{\text{step}, \mathcal{S}_{i+1}} = \max_{p_i \in \mathcal{P}} \{ t_{p_i} \}
\quad \text{where} \quad \mathcal{P} = \{p_1, p_2, \ldots, p_P\}
\end{align}
where $\mathcal{P}$ represents the set of all pipelines. 

In asymmetric parallelism, estimating the execution time of each pipeline $t_{p_i}$ is not straightforward. 
First, uneven data distribution across pipelines leads to different numbers of micro-batches for each pipeline. 
Meanwhile, unbalanced layer allocation among stages within the same pipeline directly affects the forward and backward times of each stage. As illustrated in Figure~\ref{fig:complex_pipeline}, asymmetric pipeline parallelism leads to varying forward and backward computation times across stages, resulting in numerous pipeline bubbles and making estimation challenging.
We observe that the start time of the $j$-th computation in the $i$-th stage ($t_{i, j, start}$) depends on both the end time of the previous computation in the same stage ($t_{i, j - 1, end}$) and the end time of its depended computation (often has the same micro-batch index) in the previous stage ($t_{f_1(i), f_2(j), end}$). 
Therefore, we propose a dynamic programming algorithm to simulate the pipeline time $t_{p_i}$. The general transition function during the ready phase is defined below.
\begin{align}
t_{i, j, start} &= \max(t_{i, j - 1, end}, t_{f_1(i), f_2(j), end})
\label{eq:transition} 
\end{align}
where $f_1(i)$ denotes a mapping from the current stage index $i$ to the previous stage index $i'$ depended by the current computation, for example, if the current computation is forward, $f_1(i) = i - 1$, and if it is backward, $f_1(i) = i + 1$;
and $f_2(j)$ denotes a mapping from the current computation index $j$ to the depended computation index $j'$ in the previous stage, which can be profiled from the actual execution order.

For dynamic parallelism, the state transition introduces non-negligible time overhead $t_{\mathcal{S}_i \to \mathcal{S}_{i+1}}$. 
This overhead consists of three parts: (i) the cost of searching for the optimal execution plan; (ii) the cost of transferring model weights; and (iii) the overhead of restarting training. 
It is important to note that the search can be performed in advance and overlapped with training time, while the restart overhead depends only on the training scale. 
However, the weight transfer cost varies with the execution plan and is difficult to predict.
Detailed optimization of transmission is discussed in Section~\ref{sec:restorer}.

    

\textit{Data rerouting}. 
For this policy, since no reconfiguration of the training is required, the state transition time $t_{\mathcal{S}_i \to \mathcal{S}_{i+1}}$ can be considered negligible. 
The main time overhead occurs during post-recovery training, as some nodes take on additional computation tasks of the failed nodes, resulting in a longer per-step execution time $t_\text{step}$.
Recycle optimizes through techniques such as decoupled backward propagation and staggered optimization, reducing rerouting latency via sophisticated scheduling.
To simplify the analysis, we assume that the computational tasks of the failed nodes are evenly distributed among the remaining functional nodes in the same DP group when a failure occurs.
The per-step execution time of the 1F1B pipeline in the data rerouting under a single failure can be calculated as follows.
\begin{equation}
    t_{\text{step}, \mathcal{S}_{i+1}} = (N_{pp} + N_m - 1 + \frac{N_m}{N_{dp} - 1}) \cdot (T_f + T_b) \label{eq:recycle_step_time}
\end{equation}

However, it is worth noting that as the number of failed nodes increases, the per-step time after recovery also increases.
Specifically, the per-step time $t_{\text{step}, \mathcal{S}_{i+1}}$ with $N_f$ failed nodes is related to the distribution of failed nodes across the pipeline stages.
And its calculation is shown as follows.
\begin{equation}
    t_{\text{step}, \mathcal{S}_{i+1}} = (N_{pp} + N_m - 1 + \sum_{i=0}^{N_{pp} - 1} [F_i > 0] \cdot \frac{N_m \cdot F_i}{N_{dp} - F_i} ) \cdot (T_f + T_b)\label{eq:recycle_step_time_multi}
\end{equation}
where $F_i$ is the number of failed nodes in the $i$-th stage, and the sum of $F_i$ is equal to the total number of failed nodes $N_f$.
But if any $F_i$ is larger than $N_{dp}$, the training cannot be recovered, and we must switch to dynamic parallelism.

\head{Memory Estimator}. 
To evaluate whether a given model layer splitting is appropriate, we need to determine whether OOM occurs during training. In practice, we observe that the peak memory of the $i$-th stage appears during the steady-state phase of the pipeline. 
We can decompose the peak memory for the $i$-th pipeline stage, $M_{\text{peak}, i}$, into two primary components: static memory, which includes memory for components whose size is relatively fixed during an iteration, such as model weights and optimizer states; and dynamic memory, which consists of the stored activations from the forward pass. Its peak depends on both the number of layers in the stage and the stage's position in the pipeline.
The final formula is as follows.
\begin{equation}
M_{\text{peak}, i} \approx \underbrace{N_{l_i} \cdot (m_p + m_o + m_g)}_{\text{Static Memory}} + \underbrace{(N_{pp} - i) \cdot N_{l_i} \cdot m_a}_{\text{Peak Dynamic Memory}}
\label{eq:peak_memory_decomposed}
\end{equation}
where $N_{l_i}$ is the number of layers in the $i$-th stage, $m_p$ is the average memory for the parameters of a single layer, $m_o$ is the average memory for the optimizer states of a single layer, $m_g$ is the average memory for the gradients of a single layer ($m_g = m_p$), and $m_a$ is the average memory for the activations produced by a single micro-batch for a single layer. All of the last four parameters can be estimated by profiling.


\head{Computational Complexity}. (i) Time Estimator. For symmetric parallelism and Data Rerouting, we employ analytical formulas with $O(1)$ complexity. For asymmetric parallelism, the complexity of the dynamic programming method is $O(N_{pp} \times N_{m})$. (ii) Memory Estimator only calculates the peak memory usage from profiling, and its complexity is $O(1)$.

%% file: sections/eval.tex
\section{Evaluation}
In this section, we present the evaluation of our fault-tolerant system \sysname. 
We will detail the experimental setup, including the cluster setup, baselines, and workloads. 
Following this, we will present the results obtained from our experiments, including real-world and simulation results, as well as an ablation study to understand the impacts of the core techniques of \sysname. 
Finally, we will analyze the memory usage and convergence of \sysname.

\subsection{Experimental Setup}
We evaluated \sysname with the following setup.

\head{Cluster Setup}. 
In real-world experiments, we used a cluster consisting of 32 Ascend 910B AI training accelerators~\cite{liao2021ascend,wróblewski2025parallelscanascendai}. The cluster is composed of 4 nodes, each equipped with 8 NPUs, and each NPU has 64 GB of memory. Note that \sysname's core mechanisms—execution plan search, weight transfer, and communication scheduling—are algorithmic abstractions independent of the underlying hardware. Thus, the effectiveness demonstrated on Ascend is generalizable to other platforms, such as NVIDIA GPUs.
To compare the efficiency of \sysname with other systems, we built an event-driven simulator, with details provided in Section~\ref{sec:results}.

\head{Baselines}. 
\sysname is implemented based on PyTorch 2.1.0 and Ascend chips.
We compare the performance of \sysname with the original training in the real-world cluster.
Then, we include two state-of-the-art baselines, dynamic parallelism exemplified by Oobleck and data rerouting exemplified by ReCycle, in the simulator for comparison, as they are implemented on different hardware. We exclude redundant computation methods like Bamboo and traditional checkpointing as baselines, as recent literature has demonstrated their prohibitive overhead compared to these backup-free approaches~\cite{gandhi2024recycle}.

\head{Workloads}. 
We used the 7B Llama-2 model~\cite{touvron2023llama2openfoundation} and the WikiText dataset for training. In real-world experiments, we conduct tests on 8, 16, and 32 NPUs, with parallel configurations (PP, DP, TP) including (4, 2, 1), (4, 4, 1), and (2, 2, 8). 
The batch size ranges from 16 to 64, while the micro-batch size is always set to 1. 
To ensure that \sysname can maintain stable fault-tolerant training over long periods, the total training time lasted up to 9 hours.

\subsection{Experimental Results}\label{sec:results}
The real-world results demonstrate the practical performance of \sysname in the Ascend cluster, while the simulation results provide insights into its comparative performance against other systems.

\begin{figure*}[htbp]
     \centering
     \begin{minipage}[b]{0.24\linewidth}
          \centering
          \includegraphics[width=\linewidth]{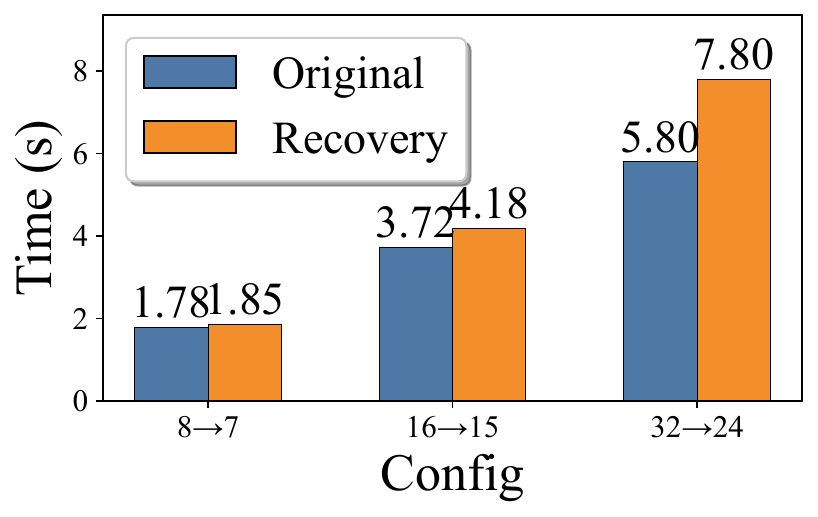}
          \caption{Real-world training.}
          \vspace{-0.2in}
          \label{fig:real-world-results}
     \end{minipage}
     \hfill
     \begin{minipage}[b]{0.22\linewidth}
          \centering
          \includegraphics[width=\linewidth]{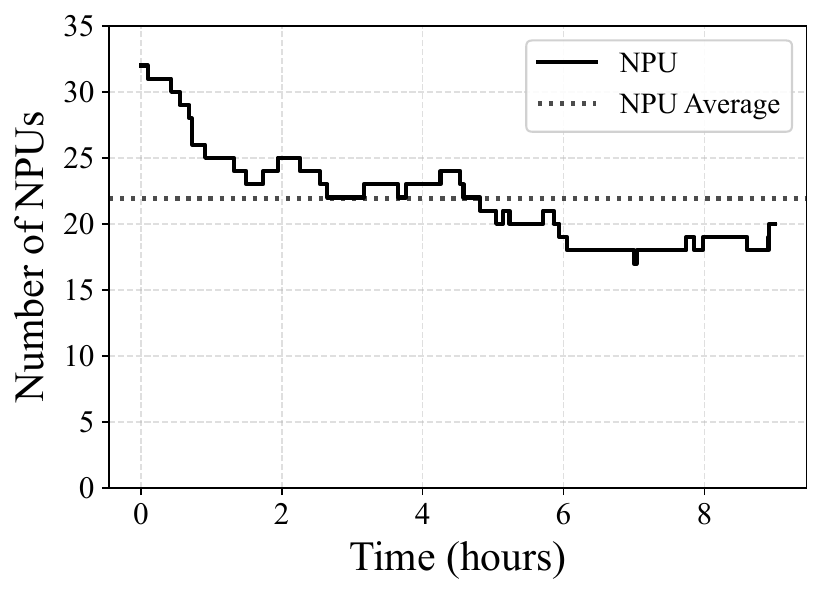}
          \caption{Num of active NPUs}
          \vspace{-0.2in}
          \label{fig:sim-subfig1}
     \end{minipage}
     \hfill
     \begin{minipage}[b]{0.22\linewidth}
          \centering
          \includegraphics[width=\linewidth]{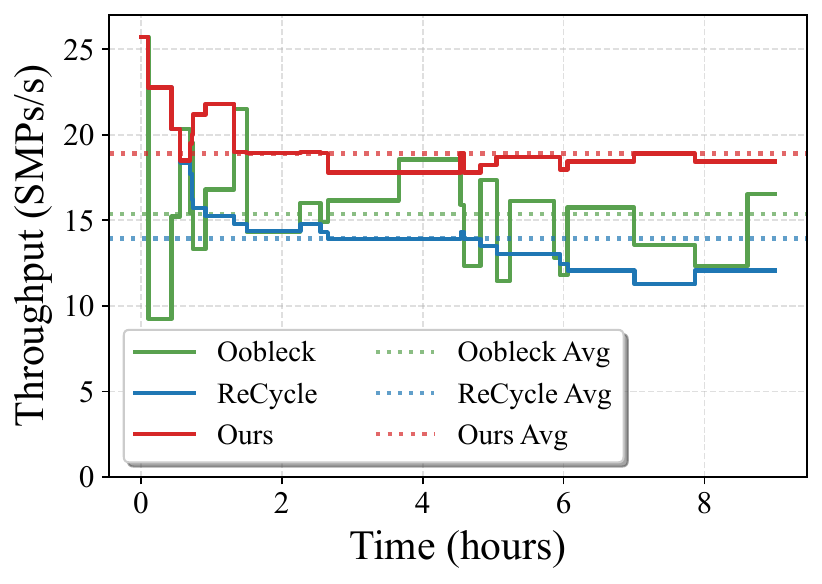}
          \caption{Training throughput}
          \vspace{-0.2in}
          \label{fig:sim-subfig2}
     \end{minipage}
     \hfill
     \begin{minipage}[b]{0.24\linewidth}
          \centering
          \includegraphics[width=\linewidth]{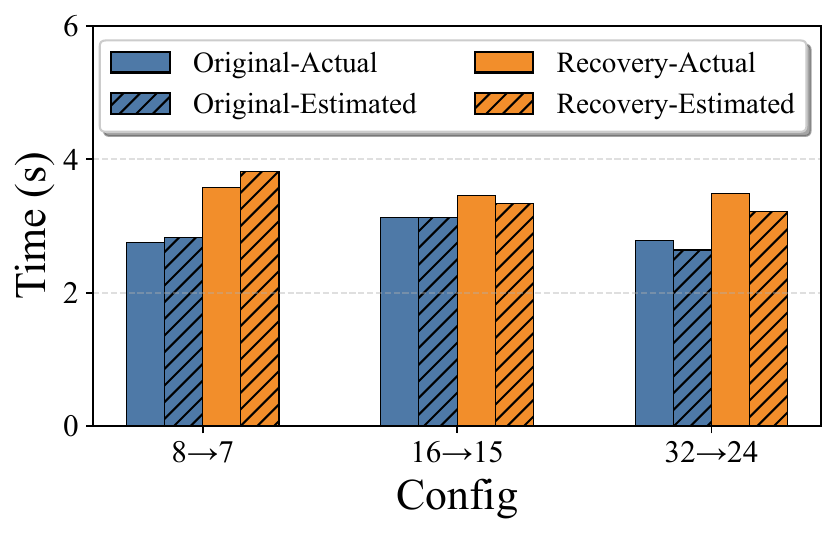}
          \caption{Estimation results}
          \vspace{-0.2in}
          \label{fig:estimation}
     \end{minipage}
\end{figure*}

\head{Real-World Results}.
Figure~\ref {fig:real-world-results} shows the performance comparison between the original training (8, 16, and 32 NPUs) and the training after failure recovery (7, 15, and 24 NPUs).
After recovery, \sysname achieves 96.22\%, 89.00\%, and 74.38\% of the original training performance, respectively. Notably, for the 32-NPU case (TP=8), a single fault leads to the loss of 8 NPUs, resulting in a larger performance gap compared to the original. 
However, compared to the theoretical maximum for 24 NPUs, \sysname reaches 99.17\% of the performance, demonstrating its ability to fully utilize available resources.

\head{Simulation Results}. 
In simulation experiments, we evaluated \sysname, ReCycle (data rerouting), and Oobleck (dynamic parallelism).
By randomly injecting failures at a specified NPU failure rate (10\% per hour), we simulated the operation of all three systems for 9 hours of training with 32 NPUs.
Figure~\ref{fig:sim-subfig1} shows the number of active nodes over time, while Figure~\ref{fig:sim-subfig2} presents the training throughput over time. 
The results show that \sysname consistently outperforms both Oobleck and ReCycle, achieving an average throughput that is 1.229$\times$ and 1.355$\times$ higher, respectively. 
While Oobleck exhibits significant throughput fluctuations, ReCycle's performance steadily degrades as the number of failures increases.

\subsection{Performance Breakdown}
To evaluate the efficiency and necessity of the core techniques in \sysname, we performed the following tests.

\head{Estimation Accuracy}. 
To verify the accuracy of the estimator, we compared the estimated time with the actual training time before and after recovery in real-world experiments under different configurations. 
As shown in Figure~\ref{fig:estimation}, the estimation error is always kept within 8.02\%, demonstrating the effectiveness of our estimation mechanism.

\begin{figure}[tbp]
     \centering
     \begin{minipage}[b]{0.48\linewidth}
          \centering
          \includegraphics[width=\linewidth]{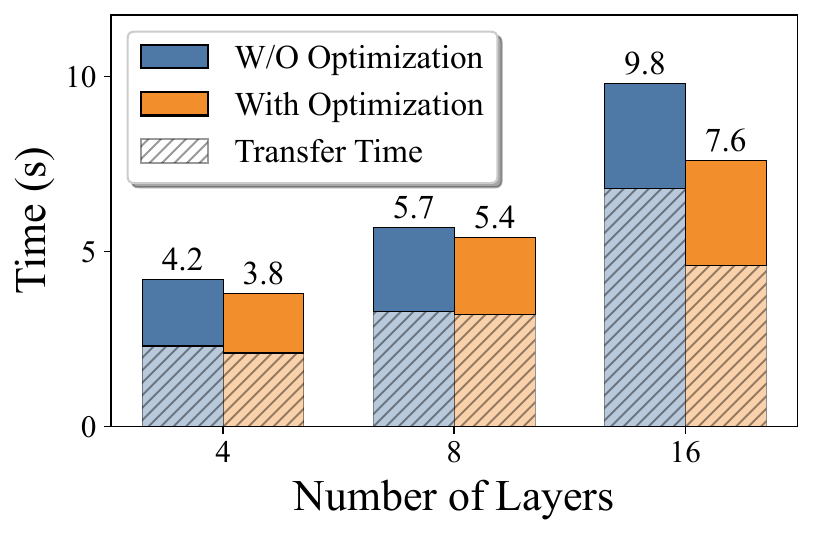}
          \caption{Weight transfer}
          \vspace{-0.2in}
          \label{fig:weight-transfer}
     \end{minipage}
     \hfill
     \begin{minipage}[b]{0.48\linewidth}
          \centering
          \includegraphics[width=\linewidth]{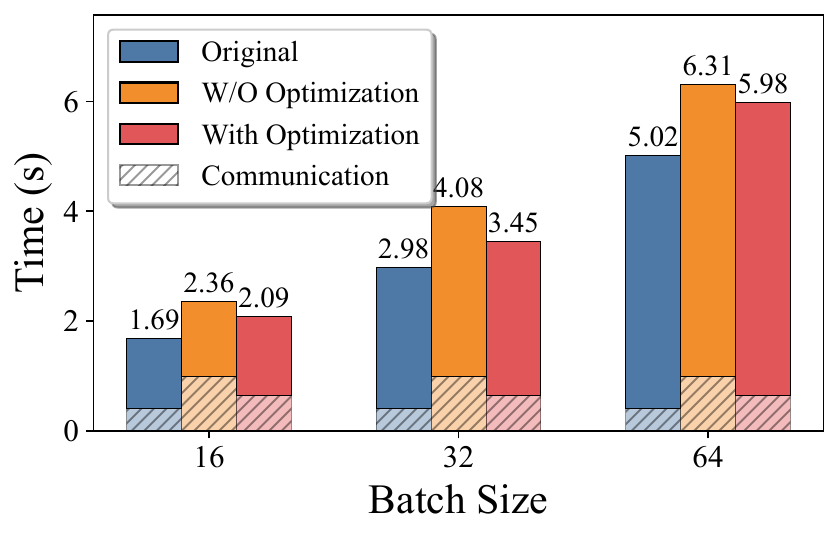}
          \caption{Asymmetric comm}
          \vspace{-0.2in}
          \label{fig:asymmetric-communication}
     \end{minipage}
\end{figure}

\head{Weight Transfer}. 
To understand the impact of weight transfer optimization on training performance, we conducted an ablation study comparing training performance with and without optimization.
Taking single-node 8-card training as an example (DP=4, PP=2, TP=1), as shown in Figure~\ref{fig:weight-transfer}, when the number of layers is relatively small (e.g., 4 or 8), the weight transfer optimization does not yield significant performance improvements. 
However, as the number of layers increases to 16, the transfer time during recovery is reduced by 32.35\%. 
Correspondingly, the total recovery time decreases from 11.2 s to 8.2 s, representing a reduction of 26.79\%, thereby enhancing overall training efficiency.

\head{Asymmetric Communication}. 
We conducted an ablation study to evaluate the impact of our asymmetric communication optimization on training performance. 
As shown in Figure~\ref{fig:asymmetric-communication}, in the same 8-card training, the batch size varies from 16 to 64, asymmetric communication optimization significantly reduces the overhead of AllReduce and other synchronizations, maintaining a reduction rate of 35.35\%. 
Although training performance decreases to some extent compared to fault-free scenarios, step time decreases by 11.44\%, 15.44\%, and 5.52\%, respectively, compared to cases without optimization. 
Notably, when the batch size is small, the optimization effect is more pronounced due to the relatively high communication ratio.

\subsection{Memory Analysis}
To ensure that \sysname can handle large-scale model training without running into OOM issues, we conducted a memory usage analysis.
Taking single-node 8-card training as an example, when an NPU failure occurs, the parallelism changes from symmetric parallelism (DP=4, PP=2) to asymmetric pipelines with length [2, 2, 3]. 
As shown in Figure~\ref{fig:memory-usage}, the peak memory per NPU in \sysname does not increase; in fact, it decreases in some cases and remains well below the device memory limit (64GB). 
This is due to changes in DP and the possible increase in the number of nodes per pipeline, resulting in fewer layers assigned to each node. 
These results demonstrate that \sysname can effectively manage memory usage during fault-tolerant training without OOMs.

\subsection{Convergence Analysis}
\begin{figure}[tbp]
\begin{minipage}[b]{0.51\linewidth}
    \centering
    \includegraphics[width=\linewidth]{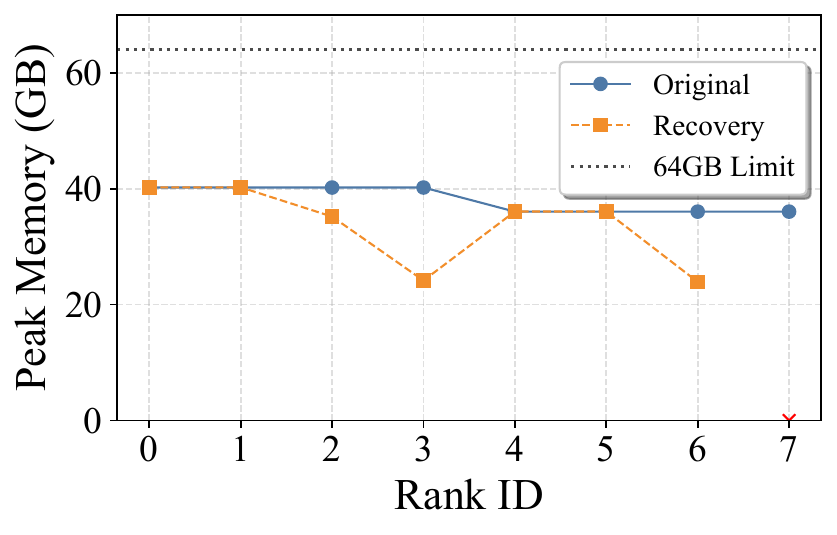}
    \caption{Memory}
    \vspace{-0.2in}
    \label{fig:memory-usage}
\end{minipage}
\hfill
\begin{minipage}[b]{0.47\linewidth}
    \centering
    \includegraphics[width=\linewidth]{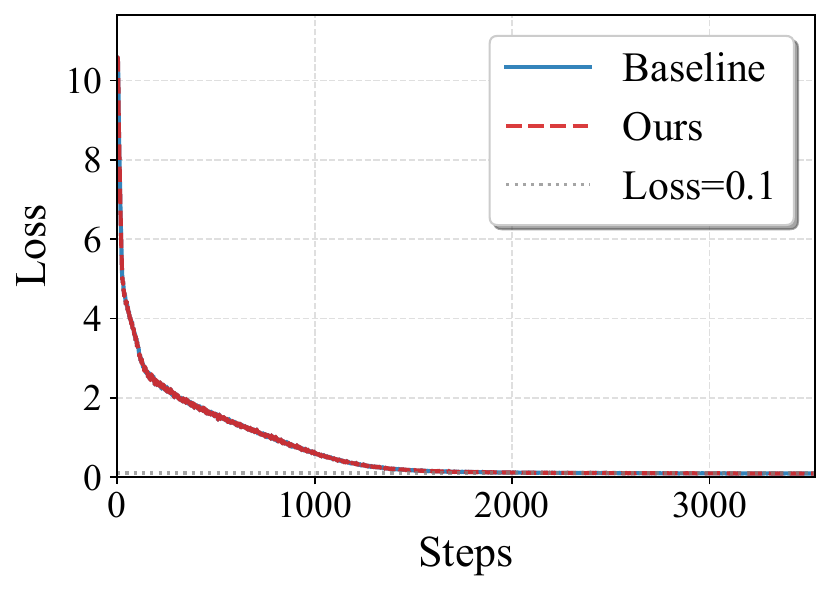}
    \caption{Convergence}
    \vspace{-0.2in}
    \label{fig:convergence-analysis}
\end{minipage}
\end{figure}

.
To verify whether \sysname will affect the model convergence, we conducted a convergence analysis on 32 Ascend NPUs by comparing our training loss with the original training. 
As shown in Figure~\ref{fig:convergence-analysis}, during training for up to 3500 steps, the training loss of \sysname remains almost identical to the original training, and both ultimately reach the convergence criterion ($\le$ 0.1). 
This indicates that \sysname does not negatively impact model convergence.

\subsection{Scalability Analysis}
To validate our theoretical scalability analysis, we measured the decision-making overhead across varying cluster sizes in simulation. For small to medium-scale clusters (up to 256 cards), the search latency is negligible ($\le 1.152$s). Even at large scales, the overhead remains acceptable (19.25s for 1024 cards and 40.48s for 2048 cards), while this search process can be pre-computed asynchronously.


%% file: sections/conclusion.tex
\section{Conclusion}\label{sec:conclusion}
This paper presents \sysname, a novel fault-tolerance system for distributed training that adaptively selects recovery policies based on real-time system states.
The core components of \sysname—the Planner, Estimator, and Restorer—enable fast and comprehensive policy search, accurate performance modeling and estimation, and low-overhead communication optimization. 
These features allow \sysname to employ the optimal execution plan, achieving more robust and efficient training compared to single-strategy approaches.
Our evaluations in both simulated and real-world environments demonstrate that \sysname significantly improves training efficiency and fault tolerance, while ensuring good memory management, model convergence, and scalability.

%% file: sections/ack.tex
\section*{Acknowledgment}
This work was supported by the Key Program of the National Natural Science Foundation of China under Grant Nos. 62325205 and 62502198, the Natural Science Foundation of Jiangsu Province under Grant Nos. BK20243053 and BK20251224, the Nanjing ``U35'' Talent Cultivation Program (No. U (2024) 001), and the Nanjing University-China Mobile Communications Group Co., Ltd. Joint Institute.

%% file: reference.bib
@article{grattafiori2024llama,
  title={The llama 3 herd of models},
  author={Grattafiori, Aaron and Dubey, Abhimanyu and Jauhri, Abhinav and others},
  journal={arXiv preprint arXiv:2407.21783},
  year={2024}
}

@inproceedings{athlur2022varuna,
  title={Varuna: scalable, low-cost training of massive deep learning models},
  author={Athlur, Sanjith and Saran, Nitika and Sivathanu, Muthian and Ramjee, Ramachandran and Kwatra, Nipun},
  booktitle={Proceedings of the Seventeenth European Conference on Computer Systems},
  pages={472--487},
  year={2022}
}

@inproceedings{thorpe2023bamboo,
  title={Bamboo: Making preemptible instances resilient for affordable training of large $\{$DNNs$\}$},
  author={Thorpe, John and Zhao, Pengzhan and Eyolfson, Jonathan and Qiao, Yifan and Jia, Zhihao and Zhang, Minjia and Netravali, Ravi and Xu, Guoqing Harry},
  booktitle={20th USENIX Symposium on Networked Systems Design and Implementation (NSDI 23)},
  pages={497--513},
  year={2023}
}

@inproceedings{jang2023oobleck,
  title={Oobleck: Resilient distributed training of large models using pipeline templates},
  author={Jang, Insu and Yang, Zhenning and Zhang, Zhen and Jin, Xin and Chowdhury, Mosharaf},
  booktitle={Proceedings of the 29th Symposium on Operating Systems Principles},
  pages={382--395},
  year={2023}
}

@inproceedings{gandhi2024recycle,
  title={ReCycle: Resilient Training of Large DNNs using Pipeline Adaptation},
  author={Gandhi, Swapnil and Zhao, Mark and Skiadopoulos, Athinagoras and Kozyrakis, Christos},
  booktitle={Proceedings of the ACM SIGOPS 30th Symposium on Operating Systems Principles},
  pages={211--228},
  year={2024}
}

@inproceedings{wagenlander2024tenplex,
  title={Tenplex: Dynamic parallelism for deep learning using parallelizable tensor collections},
  author={Wagenl{\"a}nder, Marcel and Li, Guo and Zhao, Bo and Mai, Luo and Pietzuch, Peter},
  booktitle={Proceedings of the ACM SIGOPS 30th Symposium on Operating Systems Principles},
  pages={195--210},
  year={2024}
}

@misc{qi2023zerobubble,
      title={Zero Bubble Pipeline Parallelism}, 
      author={Penghui Qi and Xinyi Wan and Guangxing Huang and Min Lin},
      year={2023},
      eprint={2401.10241},
      archivePrefix={arXiv},
      primaryClass={cs.DC},
      url={https://arxiv.org/abs/2401.10241}, 
}

@inproceedings{Deng2025Minder,
author = {Yangtao Deng and Xiang Shi and Zhuo Jiang and Xingjian Zhang and others},
title = {Minder: Faulty Machine Detection for Large-scale Distributed Model Training},
booktitle = {22nd USENIX Symposium on Networked Systems Design and Implementation (NSDI 25)},
year = {2025},
isbn = {978-1-939133-46-5},
address = {Philadelphia, PA},
pages = {505--521},
url = {https://www.usenix.org/conference/nsdi25/presentation/deng},
publisher = {USENIX Association},
month = apr
}

@inproceedings{Zhuang2023Gemini,
author = {Wang, Zhuang and Jia, Zhen and Zheng, Shuai and Zhang, Zhen and others},
title = {GEMINI: Fast Failure Recovery in Distributed Training with In-Memory Checkpoints},
year = {2023},
isbn = {9798400702297},
publisher = {Association for Computing Machinery},
address = {New York, NY, USA},
url = {https://doi.org/10.1145/3600006.3613145},
doi = {10.1145/3600006.3613145},
booktitle = {Proceedings of the 29th Symposium on Operating Systems Principles},
pages = {364–381},
numpages = {18},
location = {Koblenz, Germany},
series = {SOSP '23}
}

@misc{goyal2018accuratelargeminibatchsgd,
      title={Accurate, Large Minibatch SGD: Training ImageNet in 1 Hour}, 
      author={Priya Goyal and Piotr Dollár and Ross Girshick and Pieter Noordhuis and others},
      year={2018},
      eprint={1706.02677},
      archivePrefix={arXiv},
      primaryClass={cs.CV},
      url={https://arxiv.org/abs/1706.02677}, 
}

@misc{sergeev2018horovodfasteasydistributed,
      title={Horovod: fast and easy distributed deep learning in TensorFlow}, 
      author={Alexander Sergeev and Mike Del Balso},
      year={2018},
      eprint={1802.05799},
      archivePrefix={arXiv},
      primaryClass={cs.LG},
      url={https://arxiv.org/abs/1802.05799}, 
}

@misc{you2018imagenettrainingminutes,
      title={ImageNet Training in Minutes}, 
      author={Yang You and Zhao Zhang and Cho-Jui Hsieh and James Demmel and Kurt Keutzer},
      year={2018},
      eprint={1709.05011},
      archivePrefix={arXiv},
      primaryClass={cs.CV},
      url={https://arxiv.org/abs/1709.05011}, 
}

@misc{smith2022usingdeepspeedmegatrontrain,
      title={Using DeepSpeed and Megatron to Train Megatron-Turing NLG 530B, A Large-Scale Generative Language Model}, 
      author={Shaden Smith and Mostofa Patwary and Brandon Norick and Patrick LeGresley and others},
      year={2022},
      eprint={2201.11990},
      archivePrefix={arXiv},
      primaryClass={cs.CL},
      url={https://arxiv.org/abs/2201.11990}, 
}

@misc{rajbhandari2020zeromemoryoptimizationstraining,
      title={ZeRO: Memory Optimizations Toward Training Trillion Parameter Models}, 
      author={Samyam Rajbhandari and Jeff Rasley and Olatunji Ruwase and Yuxiong He},
      year={2020},
      eprint={1910.02054},
      archivePrefix={arXiv},
      primaryClass={cs.LG},
      url={https://arxiv.org/abs/1910.02054}, 
}

@misc{huang2019gpipeefficienttraininggiant,
      title={GPipe: Efficient Training of Giant Neural Networks using Pipeline Parallelism}, 
      author={Yanping Huang and Youlong Cheng and Ankur Bapna and others},
      year={2019},
      eprint={1811.06965},
      archivePrefix={arXiv},
      primaryClass={cs.CV},
      url={https://arxiv.org/abs/1811.06965}, 
}

@InProceedings{kim2023memorybalancedpipelineparallelism,
  title = 	 {{BP}ipe: Memory-Balanced Pipeline Parallelism for Training Large Language Models},
  author =       {Kim, Taebum and Kim, Hyoungjoo and Yu, Gyeong-In and Chun, Byung-Gon},
  booktitle = 	 {Proceedings of the 40th International Conference on Machine Learning},
  pages = 	 {16639--16653},
  year = 	 {2023},
  editor = 	 {Krause, Andreas and Brunskill, Emma and Cho, Kyunghyun and Engelhardt, Barbara and Sabato, Sivan and Scarlett, Jonathan},
  volume = 	 {202},
  series = 	 {Proceedings of Machine Learning Research},
  month = 	 {23--29 Jul},
  publisher =    {PMLR},
  url = 	 {https://proceedings.mlr.press/v202/kim23l.html}
}

@inproceedings{Li2021Chimera, series={SC ’21},
   title={Chimera: efficiently training large-scale neural networks with bidirectional pipelines},
   url={http://dx.doi.org/10.1145/3458817.3476145},
   DOI={10.1145/3458817.3476145},
   booktitle={Proceedings of the International Conference for High Performance Computing, Networking, Storage and Analysis},
   publisher={ACM},
   author={Li, Shigang and Hoefler, Torsten},
   year={2021},
   month=nov, pages={1–14},
   collection={SC ’21} }

@misc{korthikanti2022reducingactivationrecomputationlarge,
      title={Reducing Activation Recomputation in Large Transformer Models}, 
      author={Vijay Korthikanti and Jared Casper and Sangkug Lym and Lawrence McAfee and others},
      year={2022},
      eprint={2205.05198},
      archivePrefix={arXiv},
      primaryClass={cs.LG},
      url={https://arxiv.org/abs/2205.05198}, 
}

@inproceedings {Zheng2022Alpa,
author = {Lianmin Zheng and Zhuohan Li and Hao Zhang and Yonghao Zhuang and others},
title = {Alpa: Automating Inter- and {Intra-Operator} Parallelism for Distributed Deep Learning},
booktitle = {16th USENIX Symposium on Operating Systems Design and Implementation (OSDI 22)},
year = {2022},
isbn = {978-1-939133-28-1},
address = {Carlsbad, CA},
pages = {559--578},
url = {https://www.usenix.org/conference/osdi22/presentation/zheng-lianmin},
publisher = {USENIX Association},
month = jul
}

@INPROCEEDINGS{Gupta2017Failuresinlargescalesystems,
  author={Gupta, Saurabh and Patel, Tirthak and Engelmann, Christian and Tiwari, Devesh},
  booktitle={SC17: International Conference for High Performance Computing, Networking, Storage and Analysis}, 
  title={Failures in Large Scale Systems: Long-term Measurement, Analysis, and Implications}, 
  year={2017},
  volume={},
  number={},
  pages={1-12},
  doi={}}

@inproceedings {weng2022MLaaSintheWild,
author = {Qizhen Weng and Wencong Xiao and Yinghao Yu and Wei Wang and others},
title = {{MLaaS} in the Wild: Workload Analysis and Scheduling in {Large-Scale} Heterogeneous {GPU} Clusters},
booktitle = {19th USENIX Symposium on Networked Systems Design and Implementation (NSDI 22)},
year = {2022},
isbn = {978-1-939133-27-4},
address = {Renton, WA},
pages = {945--960},
url = {https://www.usenix.org/conference/nsdi22/presentation/weng},
publisher = {USENIX Association},
month = apr
}

@inproceedings {Assaf2022Check-N-Run,
author = {Assaf Eisenman and Kiran Kumar Matam and Steven Ingram and others},
title = {{Check-N-Run}: a Checkpointing System for Training Deep Learning Recommendation Models},
booktitle = {19th USENIX Symposium on Networked Systems Design and Implementation (NSDI 22)},
year = {2022},
isbn = {978-1-939133-27-4},
address = {Renton, WA},
pages = {929--943},
url = {https://www.usenix.org/conference/nsdi22/presentation/eisenman},
publisher = {USENIX Association},
month = apr
}

@misc{jiang2024megascalescalinglargelanguage,
      title={MegaScale: Scaling Large Language Model Training to More Than 10,000 GPUs}, 
      author={Ziheng Jiang and Haibin Lin and Yinmin Zhong and Qi Huang and others},
      year={2024},
      eprint={2402.15627},
      archivePrefix={arXiv},
      primaryClass={cs.LG},
      url={https://arxiv.org/abs/2402.15627}, 
}

@INPROCEEDINGS{Xie2020Elan,
  author={Xie, Lei and Zhai, Jidong and Wu, Baodong and Wang, Yuanbo and others},
  booktitle={2020 IEEE 40th International Conference on Distributed Computing Systems (ICDCS)}, 
  title={Elan: Towards Generic and Efficient Elastic Training for Deep Learning}, 
  year={2020},
  volume={},
  number={},
  pages={78-88},
  doi={10.1109/ICDCS47774.2020.00018}}

@misc{cai2025shortcutconnectedexpertparallelismaccelerating,
      title={Shortcut-connected Expert Parallelism for Accelerating Mixture-of-Experts}, 
      author={Weilin Cai and Juyong Jiang and Le Qin and others},
      year={2025},
      eprint={2404.05019},
      archivePrefix={arXiv},
      primaryClass={cs.LG},
      url={https://arxiv.org/abs/2404.05019}, 
}

@misc{qian2025epsmoeexpertpipelinescheduler,
      title={EPS-MoE: Expert Pipeline Scheduler for Cost-Efficient MoE Inference}, 
      author={Yulei Qian and Fengcun Li and Xiangyang Ji and others},
      year={2025},
      eprint={2410.12247},
      archivePrefix={arXiv},
      primaryClass={cs.CL},
      url={https://arxiv.org/abs/2410.12247}, 
}

@misc{liu2025moeparallelfoldingheterogeneous,
      title={MoE Parallel Folding: Heterogeneous Parallelism Mappings for Efficient Large-Scale MoE Model Training with Megatron Core}, 
      author={Dennis Liu and Zijie Yan and Xin Yao and Tong Liu and others},
      year={2025},
      eprint={2504.14960},
      archivePrefix={arXiv},
      primaryClass={cs.LG},
      url={https://arxiv.org/abs/2504.14960}, 
}

@article{Fedus2022Switchtransformers,
author = {Fedus, William and Zoph, Barret and Shazeer, Noam},
title = {Switch transformers: scaling to trillion parameter models with simple and efficient sparsity},
year = {2022},
issue_date = {January 2022},
publisher = {JMLR.org},
volume = {23},
number = {1},
issn = {1532-4435},
journal = {J. Mach. Learn. Res.},
month = jan,
articleno = {120},
numpages = {39}
}

@misc{jin2025megascalemoelargescalecommunicationefficienttraining,
      title={MegaScale-MoE: Large-Scale Communication-Efficient Training of Mixture-of-Experts Models in Production}, 
      author={Chao Jin and Ziheng Jiang and Zhihao Bai and others},
      year={2025},
      eprint={2505.11432},
      archivePrefix={arXiv},
      primaryClass={cs.LG},
      url={https://arxiv.org/abs/2505.11432}, 
}

@inproceedings{Li2024UnderstandingCommunicationCharacteristicsofDistributedTraining,
author = {Li, Wenxue and Liu, Xiangzhou and Li, Yuxuan and others},
title = {Understanding Communication Characteristics of Distributed Training},
year = {2024},
isbn = {9798400717581},
publisher = {Association for Computing Machinery},
address = {New York, NY, USA},
url = {https://doi.org/10.1145/3663408.3663409},
doi = {10.1145/3663408.3663409},
booktitle = {Proceedings of the 8th Asia-Pacific Workshop on Networking},
pages = {1–8},
numpages = {8},
location = {Sydney, Australia},
series = {APNet '24}
}

@INPROCEEDINGS{Moolchandani2023AMPeD,
  author={Moolchandani, Diksha and Kundu, Joyjit and others},
  booktitle={2023 IEEE International Symposium on Performance Analysis of Systems and Software (ISPASS)}, 
  title={AMPeD: An Analytical Model for Performance in Distributed Training of Transformers}, 
  year={2023},
  volume={},
  number={},
  pages={306-315},
  doi={10.1109/ISPASS57527.2023.00037}}

@misc{wróblewski2025parallelscanascendai,
      title={Parallel Scan on Ascend AI Accelerators}, 
      author={Bartłomiej Wróblewski and Gioele Gottardo and Anastasios Zouzias},
      year={2025},
      eprint={2505.15112},
      archivePrefix={arXiv},
      primaryClass={cs.DC},
      url={https://arxiv.org/abs/2505.15112}, 
}

@misc{touvron2023llama2openfoundation,
      title={Llama 2: Open Foundation and Fine-Tuned Chat Models}, 
      author={Hugo Touvron and Louis Martin and Kevin Stone and Peter Albert and others},
      year={2023},
      eprint={2307.09288},
      archivePrefix={arXiv},
      primaryClass={cs.CL},
      url={https://arxiv.org/abs/2307.09288}, 
}

@article{Kuhn2012Kuhn–Munkres,
author = {Kuhn, H.},
year = {2012},
month = {05},
pages = {},
title = {The Hungarian Method for the Assignment Problem},
volume = {2},
journal = {Naval Research Logistic Quarterly}
}

@inproceedings{liao2021ascend,
  title={Ascend: a scalable and unified architecture for ubiquitous deep neural network computing: Industry track paper},
  author={Liao, Heng and Tu, Jiajin and Xia, Jing and Liu, Hu and Zhou, Xiping and Yuan, Honghui and Hu, Yuxing},
  booktitle={2021 IEEE International Symposium on High-Performance Computer Architecture (HPCA)},
  pages={789--801},
  year={2021},
  organization={IEEE}
}

@article{achiam2023gpt,
  title={Gpt-4 technical report},
  author={Achiam, Josh and Adler, Steven and Agarwal, Sandhini and Ahmad, Lama and Akkaya, Ilge and Aleman, Florencia Leoni and Almeida, Diogo and Altenschmidt, Janko and Altman, Sam and Anadkat, Shyamal and others},
  journal={arXiv preprint arXiv:2303.08774},
  year={2023}
}

@misc{kang2025elaswaveelasticnativescalablehybridparallel,
      title={ElasWave: An Elastic-Native System for Scalable Hybrid-Parallel Training}, 
      author={Xueze Kang and Guangyu Xiang and Yuxin Wang and Hao Zhang and Yuchu Fang and Yuhang Zhou and Zhenheng Tang and Youhui Lv and Eliran Maman and Mark Wasserman and Alon Zameret and Zhipeng Bian and Shushu Chen and Zhiyou Yu and Jin Wang and Xiaoyu Wu and Yang Zheng and Chen Tian and Xiaowen Chu},
      year={2025},
      eprint={2510.00606},
      archivePrefix={arXiv},
      primaryClass={cs.DC},
      url={https://arxiv.org/abs/2510.00606}, 
}
